\documentclass[fleqn,usenatbib]{mnras}

\usepackage{newtxtext,newtxmath}

\newcommand{\Msun}{\mbox{$\mathrm{M}_{\odot}$}}
\newcommand{\Rsun}{\mbox{$\mathrm{R}_{\odot}$}}

\usepackage{stmaryrd}

\usepackage[T1]{fontenc}

\DeclareRobustCommand{\VAN}[3]{#2}
\let\VANthebibliography\thebibliography
\def\thebibliography{\DeclareRobustCommand{\VAN}[3]{##3}\VANthebibliography}

\usepackage{graphicx}	
\usepackage{amsmath}	
\usepackage{caption}
\usepackage{xcolor}

\title[The fraction of white dwarf planetary systems]{The frequency of metal-enrichment of cool helium-atmosphere white dwarfs using the DESI Early Data Release}

\author[C.\,J.\,Manser et al.]{\noindent
Christopher J. Manser,$^{1,2}$\thanks{E-mail: c.j.manser92@googlemail.com}
Boris T. G\"ansicke,$^{2}$
Paula Izquierdo,$^{2}$
Andrew Swan,${^{2}}$
Joan Najita,${^{3}}$
\newauthor
C. Rockosi,${^{4,5,6}}$
Andreia Carrillo,${^{7,8}}$
Bokyoung Kim,${^{9}}$
Siyi Xu,${^{10}}$
Arjun Dey,${^{3}}$
J.~Aguilar,${^{11}}$
S.~Ahlen,${^{12}}$
\newauthor
R.~Blum,${^{3}}$
D.~Brooks,${^{13}}$
T.~Claybaugh,${^{11}}$
K.~Dawson,${^{14}}$
A.~de la Macorra,${^{15}}$
P.~Doel,${^{13}}$
E.~Gazta\~{n}aga,${^{16,17,18}}$
\newauthor
S.~Gontcho A Gontcho,${^{11}}$
K.~Honscheid,${^{19,20,21}}$
R.~Kehoe,${^{22}}$
A.~Kremin,${^{11}}$
M.~Landriau,${^{11}}$
L.~Le~Guillou,${^{23}}$
\newauthor
Michael~E.~Levi,${^{11}}$
T.~S.~Li,${^{24}}$
A.~Meisner,${^{3}}$
R.~Miquel,${^{25,26}}$
J.~Nie,${^{27}}$
M.~Rezaie,${^{28}}$
G.~Rossi,${^{29}}$
\newauthor
E.~Sanchez,${^{30}}$
M.~Schubnell,${^{31,32}}$
G.~Tarl\'{e},${^{32}}$
B.~A.~Weaver,${^{3}}$
Z.~Zhou,${^{26}}$
H.~Zou${^{26}}$
\\
$^{1}$ Astrophysics Group, Department of Physics, Imperial College London, Prince Consort Rd, London, SW7 2AZ, UK \\
$^{2}$ Department of Physics, University of Warwick, Coventry CV4 7AL, UK \\
$^{3}$ NSF’s NOIRLab, 950 N. Cherry Avenue, Tucson, AZ 85719, USA \\
$^{4}$ Department of Astronomy and Astrophysics, University of California, Santa Cruz, 1156 High Street, Santa Cruz, CA 95065, USA \\
$^{5}$ Department of Astronomy and Astrophysics, UCO/Lick Observatory, University of California, 1156 High Street, Santa Cruz, CA 95064, USA \\
$^{6}$ University of California Observatories, 1156 High Street, Sana Cruz, CA 95065, USA \\
$^{7}$ Institute for Computational Cosmology, Department of Physics, Durham University, Durham DH1 3LE, U.K \\
$^{8}$ Centre for Extragalactic Astronomy, Department of Physics, University of Durham, South Road, Durham DH1 3LE, UK \\
$^{9}$ Institute for Astronomy, University of Edinburgh, Royal Observatory, Blackford Hill, Edinburgh EH9 3HJ, UK \\
$^{10}$ Gemini Observatory/NSF's NOIRLab, 670 N. A'ohoku Place, Hilo, Hawaii, 96720, USA \\
$^{11}$ Lawrence Berkeley National Laboratory, 1 Cyclotron Road, Berkeley, CA 94720, USA \\
$^{12}$ Physics Dept., Boston University, 590 Commonwealth Avenue, Boston, MA 02215, USA \\
$^{13}$ Department of Physics \& Astronomy, University College London, Gower Street, London, WC1E 6BT, UK \\
$^{14}$ Department of Physics and Astronomy, The University of Utah, 115 South 1400 East, Salt Lake City, UT 84112, USA \\
$^{15}$ Instituto de F\'{\i}sica, Universidad Nacional Aut\'{o}noma de M\'{e}xico,  Cd. de M\'{e}xico  C.P. 04510,  M\'{e}xico \\
$^{16}$ Institut d'Estudis Espacials de Catalunya (IEEC), 08034 Barcelona, Spain \\
$^{17}$ Institute of Cosmology \& Gravitation, University of Portsmouth, Dennis Sciama Building, Portsmouth, PO1 3FX, UK \\
$^{18}$ Institute of Space Sciences, ICE-CSIC, Campus UAB, Carrer de Can Magrans s/n, 08913 Bellaterra, Barcelona, Spain \\
$^{19}$ Center for Cosmology and AstroParticle Physics, The Ohio State University, 191 West Woodruff Avenue, Columbus, OH 43210, USA \\
$^{20}$ Department of Physics, The Ohio State University, 191 West Woodruff Avenue, Columbus, OH 43210, USA \\
$^{21}$ The Ohio State University, Columbus, 43210 OH, USA \\
$^{22}$ Department of Physics, Southern Methodist University, 3215 Daniel Avenue, Dallas, TX 75275, USA \\
$^{23}$ Sorbonne Universit\'{e}, CNRS/IN2P3, Laboratoire de Physique Nucl\'{e}aire et de Hautes Energies (LPNHE), FR-75005 Paris, France \\
$^{24}$ Department of Astronomy \& Astrophysics, University of Toronto, Toronto, ON M5S 3H4, Canada \\
$^{25}$ Instituci\'{o} Catalana de Recerca i Estudis Avan\c{c}ats, Passeig de Llu\'{\i}s Companys, 23, 08010 Barcelona, Spain \\
$^{26}$ Institut de F\'{i}sica d'Altes Energies (IFAE), The Barcelona Institute of Science and Technology, Campus UAB, 08193 Bellaterra Barcelona, Spain \\
$^{27}$ National Astronomical Observatories, Chinese Academy of Sciences, A20 Datun Rd., Chaoyang District, Beijing, 100012, P.R. China \\
$^{28}$ Department of Physics, Kansas State University, 116 Cardwell Hall, Manhattan, KS 66506, USA \\
$^{29}$ Department of Physics and Astronomy, Sejong University, Seoul, 143-747, Korea \\
$^{30}$ CIEMAT, Avenida Complutense 40, E-28040 Madrid, Spain \\
$^{31}$ Department of Physics, University of Michigan, Ann Arbor, MI 48109, USA \\
$^{32}$ University of Michigan, Ann Arbor, MI 48109, USA \\
}
\date{Accepted XXX. Received YYY; in original form ZZZ}

\pubyear{2023}

\begin{document}
\label{firstpage}
\pagerange{\pageref{firstpage}--\pageref{lastpage}}
\maketitle

\begin{abstract} 
There is overwhelming evidence that white dwarfs host planetary systems; revealed by the presence, disruption, and accretion of planetary bodies. A lower limit on the frequency of white dwarfs that host planetary material has been estimated to be $\simeq$\,25\,--\,50\,per\,cent; inferred from the ongoing or recent accretion of metals onto both hydrogen-atmosphere and warm helium-atmosphere white dwarfs. Now with the unbiased sample of white dwarfs observed by the Dark Energy Spectroscopic Instrument (DESI) survey in their Early Data Release (EDR), we have determined the frequency of metal-enrichment around cool-helium atmosphere white dwarfs as $21\pm3$\,per\,cent using a sample of 234 systems. This value is in good agreement with values determined from previous studies. With the current samples we cannot distinguish whether the frequency of planetary accretion varies with system age or host-star mass, but the DESI data release 1 will contain roughly an order of magnitude more white dwarfs than DESI EDR and will allow these parameters to be investigated.
\end{abstract}
\begin{keywords}
white dwarfs -- surveys -- planetary systems -- exoplanets
\end{keywords}



\section{Introduction}\label{sec:intro}

Planetary systems around main-sequence stars are ubiquitous. Multiple occurrence-rate studies of planetary-mass bodies show that on average there is at least one planet for every star in the Milky Way \citep{cassanetal12-1,poleskietal21-1, zhu+dong21-1}. The fraction of stars that host planetary systems is harder to constrain, with estimates that $\simeq$\,30\,per\,cent of Sun-like stars host planets with radii greater than one Earth radius, and orbital periods less than 400\,d \citep{zhuetal18-1}. As the host stars of planetary systems evolve into the giant-phase, it is expected that many of their planets orbiting at distances greater than several au will survive and continue to orbit their stars after they exhaust their fuel and evolve off the main sequence \citep{villaver+livio09-1,mustilletal14-1,veras+gaensicke15-1, lagosetal21-1, veras+hinkley21-1}.

White dwarfs are the dense, degenerate stellar remnants of main-sequence stars with initial masses $\leq 8$\,\Msun\ \citep{ibenetal97-1, dobbieetal06-1}. Their high surface gravities lead to the chemical stratification of their atmospheres, resulting in the majority of white dwarfs having either hydrogen- or helium-dominated atmospheres. Heavier elements settle out of the observable atmosphere on the diffusion timescale, which varies from a few days in hot ($\simeq$\,20\,000\,K) hydrogen-atmosphere white dwarfs, to several million years in cool ($<$\,12\,000\,K) helium-atmosphere white dwarfs \citep{koester09-1}. Thus the presence of photospheric metal-line absorption features in the spectra of white dwarfs usually\footnote{Radiative levitation can sustain the presence of heavier elements in the atmosphere of hot ($\gtrsim$\,25\,000\,K) white dwarfs \citep{chayeretal95-1}, and convection can ``dredge-up'' substantial amounts of carbon up from the core in cool ($\lesssim$\,10\,000\,K), helium-atmosphere white dwarfs \citep{pelletieretal86-1}.} reveals the recent or ongoing accretion of metals.

There is a large wealth of evidence that these metals arrive at the white dwarf predominantly from a remnant planetary system that has survived the evolution of its host star to the white dwarf phase. This is identified by (i) direct observations of disintegrating and transiting planetary-sized bodies \citep{vanderburgetal15-1, vanderburgetal20-1, gaensickeetal19-1,vanderboschetal19-1, vanderboschetal21-1, guidryetal21-1, farihietal22-1}, (ii) compact ($\simeq$\,1\,\Rsun) debris discs made of dust \citep{zuckerman+becklin87-1,jura03-1, rocchettoetal15-1, wilsonetal19-1}, and/or gas \citep{gaensickeetal06-3, manseretal20-1, dennihyetal20-2, melisetal20-1, fusilloetal21-1} produced from the complete or partial disruption of a planetary body like an asteroid or comet, (iii) the emission of x-rays from the direct bombardment of the white dwarf surface from the accretion of such a disc \citep{cunninghametal22-1}, and (iv) the resultant metal-enrichment of white dwarf atmospheres with material consistent in composition with primordial chondritic material \citep{trierweileretal23-1}, bulk-Earth \citep{zuckermanetal07-1,gaensickeetal12-1, hollandsetal17-1, hollandsetal18-1}, and more diverse bodies showing volatiles \citep{farihietal13-2, xuetal17-1,johnsonetal22-1}.

For white dwarfs, the frequency of metal-enrichment of their atmospheres can be used as a proxy for a lower limit on the occurrence rate of planetary systems that orbit them. This has been performed in several studies using hydrogen-atmosphere white dwarfs \citep{zuckermanetal03-1, koesteretal14-1, wilsonetal19-1}, as they are the most abundant type of white dwarf. \cite{zuckermanetal03-1} looked at a sample of approximately 100 hydrogen-atmosphere white dwarfs without close stellar companions, with the majority of systems having relatively cool effective temperatures ($T_{\mathrm{eff}}$\,$<$\,10\,000\,K), and identified that roughly 25\,per\,cent of them had absorption features consistent with metal-enrichment from planetary material. This result provided the first lower limit on the occurrence rate of planetary systems around white dwarfs. \cite{koesteretal14-1} utilised the \textit{Hubble Space Telescope} (HST) to look at a sample of 85 hydrogen-atmosphere white dwarfs with effective temperatures in the range 27\,000\,K\,$>$\,$T_{\mathrm{eff}}$\,$>$\,17\,000\,K. This sample of warm hydrogen-dominated white dwarfs provided a frequency of white dwarfs that have accreted planetary material ranging from $\simeq$\,25\,--\,50\,per\,cent. A more recent study using HST by \cite{wilsonetal19-1} using 143 hydrogen-atmosphere (DA) white dwarfs (which includes the sample of \citealt{koesteretal14-1}) corroborates these findings with an occurrence rate of the accretion of planetary material of $45 \pm 4$\,per\,cent. 

The frequency of planetary systems around warm helium-atmosphere white dwarfs has been investigated by \cite{zuckermanetal10-1}, who determined an occurrence rate of white dwarfs accreting planetary material of $\simeq$\,30\,per\,cent in the temperature range 19\,500\,K\,$>$\,$T_{\mathrm{eff}}$\,$>$\,13\,500\,K. This is again consistent with the occurrence rates determined from hydrogen-atmosphere white dwarfs, and the frequency of planets around main-sequence stars. This would suggest that the frequency of planetary systems around white dwarfs with hydrogen- and helium-dominated atmospheres do not differ.

The Dark Energy Spectroscopic Instrument (DESI, \citealt{DESI16-1, DESI16-2}) observes white dwarfs as part of its 5-year survey \citep{allendeprietoetal20-1}, and the Early Data Release (EDR, \citealt{desi23-1}) contains 2706 spectroscopically confirmed white dwarfs (Manser et al., submitted to MNRAS). These white dwarfs were selected using \textit{Gaia} photometry and astrometry with the selections described in \cite{cooperetal23-1}, and based on those of \cite{fusilloetal19-1,fusilloetal21-1}. The DESI EDR white dwarf sample closely follows the $G < 20$ magnitude limited sample of high-confidence white dwarfs of \cite{fusilloetal21-1}, making it far less biased than the sample of white dwarfs obtained by the Sloan Digital Sky Survey (SDSS, \citealt{gunnetal06-1, abdurroufetal22-1}) which were obtained through multiple differing targeting strategies \citep{kleinmanetal04-1}, and therefore applicable for statistical analyses.

In this paper, we use a subset of cool ($11\,500$\,K $>T_{\mathrm{eff}} > 5\,000$\,K) helium-atmosphere white dwarfs with and without signs of metal-enrichment to determine a lower limit on the occurrence rate on planetary systems around white dwarfs with cooling ages longer than $\simeq$\,400\,Myr. We give details on the white dwarfs we use for this calculation obtained from the DESI EDR white dwarf sample in Section\,\ref{sec:data}. We then calculate the frequency of white dwarf planetary systems and test the robustness of our calculations in Section\,\ref{sec:results}, and discuss our results and conclude in Section\,\ref{sec:conc}.


\section{Sample selection}\label{sec:data}

In this analysis we use the set of spectroscopically confirmed (i.e. excluding uncertain classifications, noted by ``:'') systems in the DESI EDR white dwarf sample (Manser et al., submitted to MNRAS). DESI on the Mayall 4\,m telescope at Kitt Peak National Observatory (KPNO) is a multi-object spectroscopic instrument capable of collecting fibre spectroscopy on up to $\simeq5000$ targets per pointing \citep{desi22-1}. The fibres are positioned by robot actuators and are grouped into ten petals which feed ten identical three-arm spectrographs, each spanning 3600\,--\,9824\,\AA\ at a FWHM resolution of $\simeq$\,1.8\,\AA. A full description of the DESI reduction pipeline is given by \cite{guyetal23-1}.

To determine the frequency of metal-enriched white dwarfs, we need to identify a subsample of white dwarfs with similar properties where the presence or absence of extrinsic metal can be deduced. Helium-atmosphere white dwarfs that display only the signatures of metals (excluding carbon), the DZ spectral class, are relatively easy to identify, with calcium abundances being detectable several orders of magnitude lower than their hotter counterparts that show He\,{\textsc{i}} features (DBZs) and optical absorption features that dramatically alter the emergent spectral energy distribution \citep{dufouretal07-2,koester+kepler15-1, hollandsetal17-1}. Additionally, the convection zones of DZs are deep, leading to long diffusion timescales ($>1$\,Myr) for metals to sink out of their atmospheres and thus making metal-enrichment detectable for significantly longer than their warmer counterparts \citep{koester09-1}. These attributes make DZs good proxies for determining limits on the occurrence rate of planetary systems around white dwarfs. 

\begin{figure*}
	\includegraphics[width=2\columnwidth]{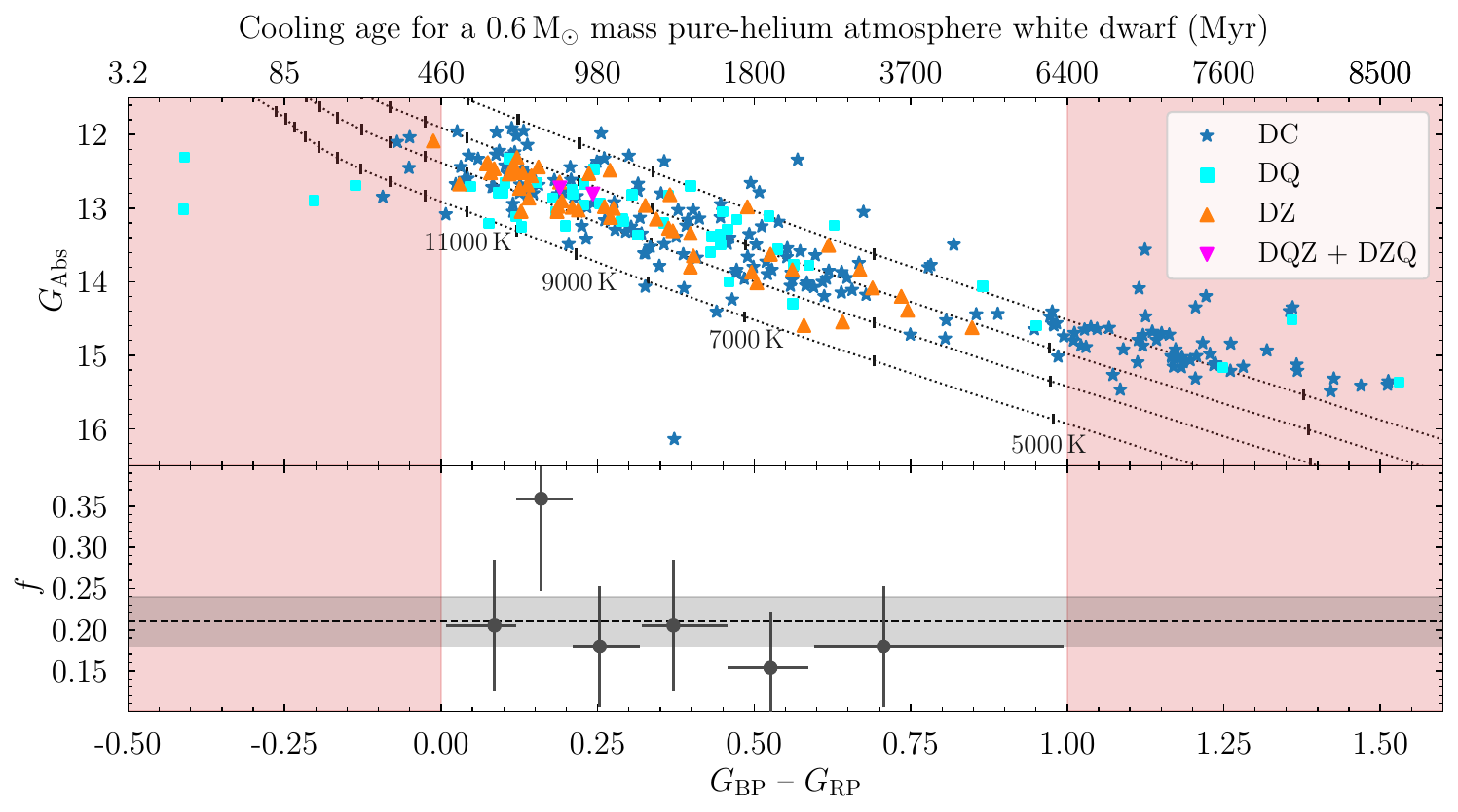}
    \caption{\textbf{Top panel:} The \textit{Gaia} Hertzsprung-Russell Diagram (HRD) showing the metal-enriched (orange and magenta triangles) and non-enriched (blue stars and cyan squares) systems used in the calculation of the frequency of planetary systems around cool, helium-atmosphere white dwarfs, $f$ (see Section\,\ref{sec:results}). Cooling tracks from \protect\cite{bedardetal20-1} for pure helium-atmosphere white dwarfs with masses from top to bottom of 0.4\,\Msun, 0.6\,\Msun, 0.8\,\Msun\ and 1.0\,\Msun\ are plotted as dashed lines on the \textit{Gaia} HRD, with vertical tabs highlighting $T_{\textrm{eff}}$ values in steps of 1000\,K. These cooling tracks are only strictly applicable to helium-atmosphere white dwarfs with no traces of metals or carbon. The red shaded regions shows the ranges excluded from the $0 < (G_\mathrm{BP} - G_\mathrm{RP}) < 1$ selection used in our $f$ calculations. One DC, WD\,J095106.36+645400.46, sits far below the cooling track with $G_\mathrm{Abs} > 16$ and is thought to have a spectrum dominated by collisionally-induced absorption \protect\citep{kilicetal20-1,bergeronetal22-1}. We still include this system in our analyses. \textbf{Bottom panel:} The fraction of metal-enriched helium-dominated white dwarfs with no trace hydrogen detectable as a function of $G_\textrm{BP}$ -- $G_\textrm{RP}$ is shown in six, equal sized bins of 39 white dwarfs. Vertical lines on the gray points represent 1\,$\sigma$ errors on the value of $f$ in the bin, while the horizontal lines represent the span of the $(G_\mathrm{BP} - G_\mathrm{RP})$ values for each bin. The horizontal dashed line and shaded region is set to the average value and uncertainty of $f = 0.21 \pm 0.03$ calculated using all 234 white dwarfs in the range $0 < (G_\mathrm{BP} - G_\mathrm{RP}) < 1$.}
    \label{fig:DZfrac}
\end{figure*} 

A DZ white dwarf can arise from the accretion of planetary material onto a featureless DC white dwarf, or a DQ showing carbon dredge-up features. The presence of metals in a white dwarf atmosphere can eradicate the spectral features of dredged-up carbon at optical wavelengths \citep{pelletieretal86-1,blouinetal22-1,hollandsetal22-1}, but at sufficiently low metal abundances, carbon dredge-up features and metal-enrichment can be observed simultaneously, such as in the DZQ\footnote{The ordering of spectral identifications is based on the strengths of the features present. For a DZQ, the non-carbon metal features (Z) are stronger than the carbon Swan bands (Q) present.}, WD\,J101453.59+411416.94, and the DQZ, WD\,J142018.91+324921.12, identified within the DESI EDR sample. This observational bias is thought to be the reason metal-enriched DQs appear to be so rare.

As the Balmer lines remain present in the optical spectrum of white dwarfs down to cooler temperatures than He\,\textsc{i} lines \citep{bergeronetal97-1}, cool helium-atmosphere white dwarfs with traces of hydrogen can have spectral types DA, DAZ, and DZA (e.g. GD362; \citealt{zuckermanetal07-1}). These are excluded from the following analysis, as without full modelling of their spectra it is difficult to distinguish them from the more common hydrogen-atmosphere white dwarfs that share those spectral types. 

Based on the above arguments, we select the metal-enriched helium-atmosphere DZs, DQZs, and DZQs, along with their non-enriched counterparts: DCs and DQs, from the DESI EDR sample (Manser et al., submitted to MNRAS). This results in a subsample of 298 white dwarfs containing 197 DCs, 50 DQs, 49 DZs, 1 DQZ, and 1 DZQ (Fig.\,\ref{fig:DZfrac}). We further restrict our sample to exclude likely contaminants using a \textit{Gaia} colour cut $0 < (G_\mathrm{BP} - G_\mathrm{RP}) < 1$. The majority of DQs bluer than $(G_\mathrm{BP} - G_\mathrm{RP}) = 0$ sit significantly below the white dwarf cooling track due to them having high masses. These systems are thought to be white dwarfs produced in binary mergers \citep{dunlap+clemens15-1,kawkaetal23-1}, and therefore do not have the same origin as the more abundant cool DQs which are likely to be the non-enriched equivalents to DZ, DZQ and DQZ white dwarfs. DCs redder than $(G_\mathrm{BP} - G_\mathrm{RP}) = 1$ are thought to be dominated by hydrogen-rich atmosphere white dwarfs as the Balmer absorption lines stop being present in the emergent spectrum \citep{caronetal23-1}, and we therefore exclude systems above this colour limit. For helium-dominated atmospheric models with no carbon or metals, the $0 < (G_\mathrm{BP} - G_\mathrm{RP}) < 1$ colour range spans the broad temperature range of $\simeq$\,$11\,500$\,K $> T_{\mathrm{eff}} > 5000$\,K \citep{bedardetal20-1}.

In the $0 < (G_\mathrm{BP} - G_\mathrm{RP}) < 1$ range the number of white dwarfs in our selection reduces to 141 DCs, 43 DQs, 48 DZs, 1 DQZ and 1 DZQ, for a total of 234 systems, while removing 1 DZ, 56 DCs, and 7 DQs. 

\section{Results}\label{sec:results}

We calculate the fraction, $f$, of cool metal-enriched helium-atmosphere white dwarfs as the number of DZ, DQZ, and DZQ white dwarfs divided by both the number of these metal-enriched systems and the number of the types they would appear as without the accretion of planetary bodies - DCs and DQs, $f = (n_\mathrm{DZ} + n_\mathrm{DQZ} + n_\mathrm{DZQ})/(n_\mathrm{DC} + n_\mathrm{DQ} + n_\mathrm{DZ} + n_\mathrm{DQZ} + n_\mathrm{DZQ})$ arriving at a value of $f$\,=\,0.21\,$\pm$\,0.03. The uncertainty on $f$ has been determined by sampling from a binomial distribution, and this method is used throughout this study. The value of $f$ obtained here is in reasonable agreement with values determined using cool hydrogen-atmosphere white dwarfs \citep{zuckermanetal03-1}, but significantly lower than those determined for warmer hydrogen- and helium-atmosphere white dwarfs \citep{zuckermanetal10-1, koesteretal14-1, wilsonetal19-1}.

We investigate the dependence of $f$ on $(G_\mathrm{BP} - G_\mathrm{RP})$ in Fig.\,\ref{fig:DZfrac}, where we calculate $f$ in six equal-sized bins of 39 white dwarfs across the $0 < (G_\mathrm{BP} - G_\mathrm{RP}) < 1$. $f$ varies between $\simeq$0.15\,--\,0.36 throughout the selected colour range of $0 < (G_\mathrm{BP} - G_\mathrm{RP}) < 1$, with five out of six bins in excellent agreement with the constant value of $f = 0.21\pm0.03$, which would suggest there is no detectable trend between $f$ and $(G_\mathrm{BP} - G_\mathrm{RP})$.

\begin{table*} 
\centering
\caption{The fraction of white dwarfs that have accreted planetary material, $f$ from various studies. The sample size column gives both the total sample size, with the number of metal-enriched white dwarf in brackets. Cooling ages, $\tau$, are estimated based on the temperature range of the sample and white dwarf evolutionary models of \protect\cite{bedardetal20-1}. \label{t-WD_planet_fractions}}
\begin{tabular}{lllll} 
\hline
Sample                                                 & WD type                    & Sample size   & Fraction ($f$)  & Cooling age ($\tau$) [Myr]    \\
\hline
This work, $0 < (G_\mathrm{BP} - G_\mathrm{RP}) < 1$   & helium-atmosphere (DC + DQ)  & 234 (50)      & $0.21 \pm 0.03$ & $\sim$\,300\,-\,10\,000   \\
This work, $0 < (G_\mathrm{BP} - G_\mathrm{RP}) < 1$   & helium-atmosphere (DC)       & 189 (48)      & $0.25 \pm 0.03$ & $\sim$\,300\,-\,10\,000   \\
This work                                              & DC + DQ $^\mathrm{a}$        & 298 (51)      & $0.17 \pm 0.02$ & $\sim$\,100\,-\,10\,000   \\
This work                                              & DC $^\mathrm{a}$             & 246 (49)      & $0.20 \pm 0.03$ & $\sim$\,100\,-\,10\,000   \\
\cite{hollandsetal18-2}                                & DC + DQ $^\mathrm{a}$        & 51 (9)        & $0.18 \pm 0.05$ & $\sim$\,100\,-\,10\,000   \\
\cite{hollandsetal18-2}                                & DC $^\mathrm{a}$             & 38 (9)        & $0.24 \pm 0.07$ & $\sim$\,100\,-\,10\,000   \\
\cite{zuckermanetal10-1}                               & helium-atmosphere (DB + DBA) & 54 (16)       & 0.30            & $\sim$\,100\,-\,300 \\
\cite{zuckermanetal03-1}                               & hydrogen-atmosphere (DA)     & $\simeq$\,80  ($\simeq$\,20) & $\simeq$\,0.25  & $\sim$\,600\,-\,10\,000  \\
\cite{koesteretal14-1}\,$^\mathrm{b}$ actively accreting & hydrogen-atmosphere (DA)     & 85 (23)       & 0.27            & $\sim$\,10\,-\,300  \\
\cite{koesteretal14-1}\,$^\mathrm{b}$ metals present     & hydrogen-atmosphere (DA)     & 85 (48)       & 0.56            & $\sim$\,10\,-\,300  \\
\cite{wilsonetal19-1}\,$^\mathrm{b}$                     & hydrogen-atmosphere (DA)     & 143 (65)      & $0.45 \pm 0.04$ & $\sim$\,10\,-\,300  \\ 
\hline
\multicolumn{5}{|l|}{$^\mathrm{a}$ No colour cuts have been placed and it is likely that the sample includes hydrogen-atmosphere DCs and high-mass binary-merger DQ white dwarfs.} \\
\multicolumn{5}{|l|}{$^\mathrm{b}$ Sample observed exclusively with \textit{HST} rather than ground-based observatories.} \\
\end{tabular}
\end{table*}

Our calculation also relies on the assumption that the colour cut $0 < (G_\mathrm{BP} - G_\mathrm{RP}) < 1$ we make cleanly selects a sample of cool-helium dominated white dwarfs and rejects contaminant systems. Relaxing this cut and using all 298 white dwarfs with spectral types of DC, DQ, DZ, DZQ and DQZ results in a value of $f$\,=\,0.17\,$\pm$\,0.03, and can be taken as a conservative lower-limit on the fraction of metal-enriched cool helium-atmosphere white dwarfs. This value is also consistent with our previous estimate and a larger sample size is needed to identify a significant difference between the two values of $f$ obtained so far.

We also compare our values of $f$ to those obtainable using the volume limited 20\,pc sample of white dwarfs from \cite{hollandsetal18-2}, which contains 29 DC and DCP white dwarfs, 13 DQ, DQPec and DQPecP white dwarfs, and 9 DZ white dwarfs, where `P' and `Pec' identify the detection of polarisation due to magnetism and peculiar/unknown features respectively. The resulting value of $f$\,=\,0.18\,$\pm$\,0.06 obtained from the entire 20\,pc sample is in excellent agreement with the value of $f$ we obtain from the DESI EDR sample. Finally, there are some arguments that all DQ white dwarfs are the product of binary merger events \citep{farihietal22-2}, and as such should not be considered to be in the same population as the standard, isolated cool DC white dwarfs. If this is the case, we also calculate $f$ values excluding DQs, DQZs and DZQs, which are given in Table\,\ref{t-WD_planet_fractions}, and are still in good agreement with $f$ values including carbon-line DQ white dwarfs. 

These tests demonstrate the robustness of our assumptions with the available data sets and that the frequency we determine here is representative for planetary accretion at cool helium-dominated white dwarfs. We present all the occurrence rates of planetary accretion at white dwarfs we discuss in Table\,\ref{t-WD_planet_fractions}.

\section{Discussion and Conclusions}\label{sec:conc}

Helium-atmosphere white dwarfs in the temperature range $\simeq$\,$11\,500$\,K $> T_{\mathrm{eff}} > 5000$\,K have cooling ages, $\tau$, ranging from 460\,Myr\,$< \tau <$\,6.4\,Gyr for a 0.6\,\Msun white dwarf \citep{bedardetal20-1}. Our sample therefore probes a significantly older range of observed white dwarfs compared with the relatively young samples of \cite{koesteretal14-1}, \cite{wilsonetal19-1}, and \cite{zuckermanetal10-1} which span ages in the range 9\,Myr\,$< \tau <$\,300\,Myr. The sample of hydrogen-dominated white dwarfs investigated by \cite{zuckermanetal03-1} extends over a temperature and age range consistent with our sample.

The occurrence rates of planetary systems around white dwarfs obtained by \cite{wilsonetal19-1} and \cite{koesteretal14-1} are at least as high as $0.45 \pm 0.04$ and $0.56 \pm 0.10$ respectively, significantly higher than the value of $0.21 \pm 0.03$ we derive here. Conversely the occurrence of planetary systems in the samples of \cite{zuckermanetal03-1} and \cite{zuckermanetal10-1} reduces to $\simeq$\,0.25\,-\,0.30, slightly higher than but still in agreement with the value we obtain here. While this could be suggestive of the depletion of planetary material around white dwarfs as a function of cooling age, \cite{koesteretal14-1} note that this discrepancy is likely due to the high resolution, high signal-to-noise ratio, and ultraviolet wavelength-range obtained by the HST Cosmic Origins Spectrograph compared with the ground-based observations of \cite{zuckermanetal03-1,zuckermanetal10-1} and the DESI EDR sample. Higher-resolution and deeper follow-up of the full DESI EDR sample of cool helium-dominated white dwarfs could reveal additional metal-enriched systems undetected by DESI or identify contaminant systems (e.g. white dwarfs with weak or magnetically-split hydrogen lines), both of which would increase the lower limit of $f$ determined here. 

\cite{hollandsetal18-1} found a depletion in the accretion rates on an $e$-folding timescale of $\simeq$\,1\,Gyr around cool DZ white dwarfs in the cooling-age range 1\,Gyr\,$< \tau <$\,8\,Gyr. However, this result has been disputed by \citep{blouin+xu22-1} who used updated white dwarf models along with \textit{Gaia} parallaxes and identified at best only a factor ten drop off in accretion rates over the range of 1\,Gyr\,$< \tau <$\,8\,Gyr. While these studies suggest a potential depletion or constant rate of planetary material accreted by white dwarfs over several billion years, they cannot be used to directly probe changes to the occurrence rate of planetary systems around white dwarfs over the same time-frame. The detection sensitivity of planetary material in the atmospheres of white dwarfs is dependent on both the amount of material that is being delivered to and accreted by white dwarfs in addition to the frequency of planetary systems around them. Forward-modelling of the samples of white dwarfs in Table\,\ref{t-WD_planet_fractions} would allow the evolution of the accretion and frequency of planetary systems to be constrained based on the current understanding of model atmospheres \citep{baueretal18-1,cunninghametal19-1}, and inform future observing campaigns. 

While we ignore the presence of hydrogen in our analysis, there is a significant link between the presence of hydrogen in a helium-atmosphere white dwarf, and the presence of metals \citep{fusilloetal17-1}. This correlation in the \textit{presence} of hydrogen and metals has been explained as either: (i) the accretion of planetary debris being linked to the accretion of hydrogen in the form of water \citep{fusilloetal17-1}, or the accretion of giant planet atmospheres \citep{schreiberetal19-1}, or (ii) that helium-atmosphere white dwarfs born without detectable trace hydrogen have a significantly reduced planetary system occurrence rate \citep{bedardetal23-1}. Investigating the link between the presence of hydrogen and metals is beyond the scope of this paper, but the frequencies we calculate here for cool helium-atmosphere white dwarfs along with those in Table.\,\ref{t-WD_planet_fractions} will allow the degeneracy between the two hypotheses listed to be lifted.

The DESI survey is well into its five year observing program, and the first data release (DR1, currently internal) contains spectroscopy of over 47\,000 white dwarf candidates \citep{manseretal23-1}. Based on the DESI EDR sample of spectroscopically confirmed white dwarfs, it is expected that the DR1 sample with contain over 2\,000 helium-dominated white dwarfs, and roughly 500 metal-enriched white dwarfs in the colour range $0 < (G_\mathrm{BP} - G_\mathrm{RP}) < 1$. Such a sample would allow the frequency of planetary systems around white dwarfs to be probed in much greater detail, such as whether there is a drop-off in the occurrence of planetary systems at late cooling ages. Additionally, this sample would allow other studies to be performed, such as an independent test of observed accretion rates with white dwarf cooling ages, or differences in accreted material as a function of white dwarf mass and therefore host-star progenitor mass, studies of which are currently limited \citep{johnsonetal10-1,zhu+dong21-1}.

\section*{Acknowledgements}

The authors acknowledge financial support from Imperial College London through an Imperial College Research Fellowship grant awarded to CJM. This project has received funding from the European Research Council (ERC) under the European Union’s Horizon 2020 research and innovation programme (Grant agreement No. 101020057). SX is supported by NOIRLab, which is managed by the Association of Universities for Research in Astronomy (AURA) under a cooperative agreement with the National Science Foundation.
This material is based upon work supported by the U.S. Department of Energy (DOE), Office of Science, Office of High-Energy Physics, under Contract No. DE–AC02–05CH11231, and by the National Energy Research Scientific Computing Center, a DOE Office of Science User Facility under the same contract. Additional support for DESI was provided by the U.S. National Science Foundation (NSF), Division of Astronomical Sciences under Contract No. AST-0950945 to the NSF’s National Optical-Infrared Astronomy Research Laboratory; the Science and Technology Facilities Council of the United Kingdom; the Gordon and Betty Moore Foundation; the Heising-Simons Foundation; the French Alternative Energies and Atomic Energy Commission (CEA); the National Council of Science and Technology of Mexico (CONACYT); the Ministry of Science and Innovation of Spain (MICINN), and by the DESI Member Institutions: \url{https://www.desi.lbl.gov/collaborating-institutions}. Any opinions, findings, and conclusions or recommendations expressed in this material are those of the author(s) and do not necessarily reflect the views of the U. S. National Science Foundation, the U. S. Department of Energy, or any of the listed funding agencies.

The authors are honored to be permitted to conduct scientific research on Iolkam Du’ag (Kitt Peak), a mountain with particular significance to the Tohono O’odham Nation.

This work has made use of data from the European Space Agency (ESA) mission
{\it Gaia} (\url{https://www.cosmos.esa.int/gaia}), processed by the {\it Gaia}
Data Processing and Analysis Consortium (DPAC,
\url{https://www.cosmos.esa.int/web/gaia/dpac/consortium}). Funding for the DPAC
has been provided by national institutions, in particular the institutions
participating in the {\it Gaia} Multilateral Agreement.

\section*{Data availability}
The data presented here are all available from the public archives of DESI. Data used to produce the figures presented in this paper are available under \texttt{DOI: 10.5281/zenodo.10226244} and can be found here \url{https://zenodo.org/records/10226244}.



\bibliographystyle{mnras}
\bibliography{aamnem99,aabib_planet_pap,Swan}

\begin{thebibliography}{}
\makeatletter
\relax
\def\mn@urlcharsother{\let\do\@makeother \do\$\do\&\do\#\do\^\do\_\do\%\do\~}
\def\mn@doi{\begingroup\mn@urlcharsother \@ifnextchar [ {\mn@doi@} {\mn@doi@[]}}
\def\mn@doi@[#1]#2{\def\@tempa{#1}\ifx\@tempa\@empty \href {http://dx.doi.org/#2} {doi:#2}\else \href {http://dx.doi.org/#2} {#1}\fi \endgroup}
\def\mn@eprint#1#2{\mn@eprint@#1:#2::\@nil}
\def\mn@eprint@arXiv#1{\href {http://arxiv.org/abs/#1} {{\tt arXiv:#1}}}
\def\mn@eprint@dblp#1{\href {http://dblp.uni-trier.de/rec/bibtex/#1.xml} {dblp:#1}}
\def\mn@eprint@#1:#2:#3:#4\@nil{\def\@tempa {#1}\def\@tempb {#2}\def\@tempc {#3}\ifx \@tempc \@empty \let \@tempc \@tempb \let \@tempb \@tempa \fi \ifx \@tempb \@empty \def\@tempb {arXiv}\fi \@ifundefined {mn@eprint@\@tempb}{\@tempb:\@tempc}{\expandafter \expandafter \csname mn@eprint@\@tempb\endcsname \expandafter{\@tempc}}}

\bibitem[\protect\citeauthoryear{{Abdurro'uf} et~al.,}{{Abdurro'uf} et~al.}{2022}]{abdurroufetal22-1}
{Abdurro'uf} et~al., 2022, \mn@doi [\apjs] {10.3847/1538-4365/ac4414}, \href {https://ui.adsabs.harvard.edu/abs/2022ApJS..259...35A} {259, 35}

\bibitem[\protect\citeauthoryear{{Allende Prieto} et~al.,}{{Allende Prieto} et~al.}{2020}]{allendeprietoetal20-1}
{Allende Prieto} C.,  et~al., 2020, \mn@doi [Research Notes of the American Astronomical Society] {10.3847/2515-5172/abc1dc}, \href {https://ui.adsabs.harvard.edu/abs/2020RNAAS...4..188A} {4, 188}

\bibitem[\protect\citeauthoryear{{Bauer} \& {Bildsten}}{{Bauer} \& {Bildsten}}{2018}]{baueretal18-1}
{Bauer} E.~B.,  {Bildsten} L.,  2018, \mn@doi [\apjl] {10.3847/2041-8213/aac492}, \href {https://ui.adsabs.harvard.edu/abs/2018ApJ...859L..19B} {859, L19}

\bibitem[\protect\citeauthoryear{{B{\'e}dard}, {Bergeron}, {Brassard}  \& {Fontaine}}{{B{\'e}dard} et~al.}{2020}]{bedardetal20-1}
{B{\'e}dard} A.,  {Bergeron} P.,  {Brassard} P.,   {Fontaine} G.,  2020, \mn@doi [\apj] {10.3847/1538-4357/abafbe}, \href {https://ui.adsabs.harvard.edu/abs/2020ApJ...901...93B} {901, 93}

\bibitem[\protect\citeauthoryear{{B{\'e}dard}, {Bergeron}  \& {Brassard}}{{B{\'e}dard} et~al.}{2023}]{bedardetal23-1}
{B{\'e}dard} A.,  {Bergeron} P.,   {Brassard} P.,  2023, \mn@doi [\apj] {10.3847/1538-4357/acbb62}, \href {https://ui.adsabs.harvard.edu/abs/2023ApJ...946...24B} {946, 24}

\bibitem[\protect\citeauthoryear{{Bergeron}, {Ruiz}  \& {Leggett}}{{Bergeron} et~al.}{1997}]{bergeronetal97-1}
{Bergeron} P.,  {Ruiz} M.~T.,   {Leggett} S.~K.,  1997, \mn@doi [ApJS] {10.1086/312955}, \href {1997ApJS..108..339B} {108, 339}

\bibitem[\protect\citeauthoryear{{Bergeron}, {Kilic}, {Blouin}, {B{\'e}dard}, {Leggett}  \& {Brown}}{{Bergeron} et~al.}{2022}]{bergeronetal22-1}
{Bergeron} P.,  {Kilic} M.,  {Blouin} S.,  {B{\'e}dard} A.,  {Leggett} S.~K.,   {Brown} W.~R.,  2022, \mn@doi [\apj] {10.3847/1538-4357/ac76c7}, \href {https://ui.adsabs.harvard.edu/abs/2022ApJ...934...36B} {934, 36}

\bibitem[\protect\citeauthoryear{{Blouin}}{{Blouin}}{2022}]{blouinetal22-1}
{Blouin} S.,  2022, \mn@doi [\aap] {10.1051/0004-6361/202244944}, \href {https://ui.adsabs.harvard.edu/abs/2022A&A...666L...7B} {666, L7}

\bibitem[\protect\citeauthoryear{{Blouin} \& {Xu}}{{Blouin} \& {Xu}}{2022}]{blouin+xu22-1}
{Blouin} S.,  {Xu} S.,  2022, \mn@doi [\mnras] {10.1093/mnras/stab3446}, \href {https://ui.adsabs.harvard.edu/abs/2022MNRAS.510.1059B} {510, 1059}

\bibitem[\protect\citeauthoryear{{Caron}, {Bergeron}, {Blouin}  \& {Leggett}}{{Caron} et~al.}{2023}]{caronetal23-1}
{Caron} A.,  {Bergeron} P.,  {Blouin} S.,   {Leggett} S.~K.,  2023, \mn@doi [\mnras] {10.1093/mnras/stac3733}, \href {https://ui.adsabs.harvard.edu/abs/2023MNRAS.519.4529C} {519, 4529}

\bibitem[\protect\citeauthoryear{{Cassan} et~al.,}{{Cassan} et~al.}{2012}]{cassanetal12-1}
{Cassan} A.,  et~al., 2012, \mn@doi [Nat] {10.1038/nature10684}, \href {2012Natur.481..167C} {481, 167}

\bibitem[\protect\citeauthoryear{{Chayer}, {Fontaine}  \& {Wesemael}}{{Chayer} et~al.}{1995}]{chayeretal95-1}
{Chayer} P.,  {Fontaine} G.,   {Wesemael} F.,  1995, ApJS, \href {1995ApJS...99..189C} {99, 189}

\bibitem[\protect\citeauthoryear{{Cooper} et~al.,}{{Cooper} et~al.}{2023}]{cooperetal23-1}
{Cooper} A.~P.,  et~al., 2023, \mn@doi [\apj] {10.3847/1538-4357/acb3c0}, \href {https://ui.adsabs.harvard.edu/abs/2023ApJ...947...37C} {947, 37}

\bibitem[\protect\citeauthoryear{{Cunningham}, {Tremblay}, {Freytag}, {Ludwig}  \& {Koester}}{{Cunningham} et~al.}{2019}]{cunninghametal19-1}
{Cunningham} T.,  {Tremblay} P.-E.,  {Freytag} B.,  {Ludwig} H.-G.,   {Koester} D.,  2019, \mn@doi [\mnras] {10.1093/mnras/stz1759}, \href {https://ui.adsabs.harvard.edu/abs/2019MNRAS.488.2503C} {488, 2503}

\bibitem[\protect\citeauthoryear{{Cunningham}, {Wheatley}, {Tremblay}, {G{\"a}nsicke}, {King}, {Toloza}  \& {Veras}}{{Cunningham} et~al.}{2022}]{cunninghametal22-1}
{Cunningham} T.,  {Wheatley} P.~J.,  {Tremblay} P.-E.,  {G{\"a}nsicke} B.~T.,  {King} G.~W.,  {Toloza} O.,   {Veras} D.,  2022, \mn@doi [\nat] {10.1038/s41586-021-04300-w}, \href {https://ui.adsabs.harvard.edu/abs/2022Natur.602..219C} {602, 219}

\bibitem[\protect\citeauthoryear{{DESI Collaboration} et~al.,}{{DESI Collaboration} et~al.}{2016a}]{DESI16-1}
{DESI Collaboration} et~al., 2016a, preprint, \href {http://ukads.nottingham.ac.uk/abs/2016arXiv161100036D} {} (\mn@eprint {arXiv} {1611.00036})

\bibitem[\protect\citeauthoryear{{DESI Collaboration} et~al.,}{{DESI Collaboration} et~al.}{2016b}]{DESI16-2}
{DESI Collaboration} et~al., 2016b, preprint, \href {http://ukads.nottingham.ac.uk/abs/2016arXiv161100037D} {} (\mn@eprint {arXiv} {1611.00037})

\bibitem[\protect\citeauthoryear{{DESI Collaboration} et~al.,}{{DESI Collaboration} et~al.}{2022}]{desi22-1}
{DESI Collaboration} et~al., 2022, \mn@doi [\aj] {10.3847/1538-3881/ac882b}, \href {https://ui.adsabs.harvard.edu/abs/2022AJ....164..207D} {164, 207}

\bibitem[\protect\citeauthoryear{{DESI Collaboration} et~al.,}{{DESI Collaboration} et~al.}{2023}]{desi23-1}
{DESI Collaboration} et~al., 2023, \mn@doi [arXiv e-prints] {10.48550/arXiv.2306.06308}, \href {https://ui.adsabs.harvard.edu/abs/2023arXiv230606308D} {p. arXiv:2306.06308}

\bibitem[\protect\citeauthoryear{{Dennihy} et~al.,}{{Dennihy} et~al.}{2020}]{dennihyetal20-2}
{Dennihy} E.,  et~al., 2020, \mn@doi [\apj] {10.3847/1538-4357/abc339}, \href {https://ui.adsabs.harvard.edu/abs/2020ApJ...905....5D} {905, 5}

\bibitem[\protect\citeauthoryear{{Dobbie} et~al.,}{{Dobbie} et~al.}{2006}]{dobbieetal06-1}
{Dobbie} P.~D.,  et~al., 2006, \mn@doi [MNRAS] {10.1111/j.1365-2966.2006.10311.x}, \href {2006MNRAS.369..383D} {369, 383}

\bibitem[\protect\citeauthoryear{{Dufour} et~al.,}{{Dufour} et~al.}{2007}]{dufouretal07-2}
{Dufour} P.,  et~al., 2007, \mn@doi [ApJ] {10.1086/518468}, \href {2007ApJ...663.1291D} {663, 1291}

\bibitem[\protect\citeauthoryear{{Dunlap} \& {Clemens}}{{Dunlap} \& {Clemens}}{2015}]{dunlap+clemens15-1}
{Dunlap} B.~H.,  {Clemens} J.~C.,  2015, in {Dufour} P.,  {Bergeron} P.,   {Fontaine} G.,  eds,  Astronomical Society of the Pacific Conference Series Vol. 493, 19th European Workshop on White Dwarfs. p.~547

\bibitem[\protect\citeauthoryear{{Farihi}, {G{\"a}nsicke}  \& {Koester}}{{Farihi} et~al.}{2013}]{farihietal13-2}
{Farihi} J.,  {G{\"a}nsicke} B.~T.,   {Koester} D.,  2013, \mn@doi [Science] {10.1126/science.1239447}, \href {2013Sci...342..218F} {342, 218}

\bibitem[\protect\citeauthoryear{{Farihi}, {Dufour}  \& {Wilson}}{{Farihi} et~al.}{2022a}]{farihietal22-2}
{Farihi} J.,  {Dufour} P.,   {Wilson} T.~G.,  2022a, \mn@doi [arXiv e-prints] {10.48550/arXiv.2208.05990}, \href {https://ui.adsabs.harvard.edu/abs/2022arXiv220805990F} {p. arXiv:2208.05990}

\bibitem[\protect\citeauthoryear{{Farihi} et~al.,}{{Farihi} et~al.}{2022b}]{farihietal22-1}
{Farihi} J.,  et~al., 2022b, \mn@doi [\mnras] {10.1093/mnras/stab3475}, \href {https://ui.adsabs.harvard.edu/abs/2022MNRAS.511.1647F} {511, 1647}

\bibitem[\protect\citeauthoryear{{G{\"a}nsicke}, {Marsh}, {Southworth}  \& {Rebassa-Mansergas}}{{G{\"a}nsicke} et~al.}{2006}]{gaensickeetal06-3}
{G{\"a}nsicke} B.~T.,  {Marsh} T.~R.,  {Southworth} J.,   {Rebassa-Mansergas} A.,  2006, \mn@doi [Science] {10.1126/science.1135033}, \href {2006Sci...314.1908G} {314, 1908}

\bibitem[\protect\citeauthoryear{{G{\"a}nsicke}, {Koester}, {Farihi}, {Girven}, {Parsons}  \& {Breedt}}{{G{\"a}nsicke} et~al.}{2012}]{gaensickeetal12-1}
{G{\"a}nsicke} B.~T.,  {Koester} D.,  {Farihi} J.,  {Girven} J.,  {Parsons} S.~G.,   {Breedt} E.,  2012, \mn@doi [MNRAS] {10.1111/j.1365-2966.2012.21201.x}, \href {2012MNRAS.424..333G} {424, 333}

\bibitem[\protect\citeauthoryear{{G{\"a}nsicke}, {Schreiber}, {Toloza}, {Fusillo}, {Koester}  \& {Manser}}{{G{\"a}nsicke} et~al.}{2019}]{gaensickeetal19-1}
{G{\"a}nsicke} B.~T.,  {Schreiber} M.~R.,  {Toloza} O.,  {Fusillo} N. P.~G.,  {Koester} D.,   {Manser} C.~J.,  2019, \mn@doi [\nat] {10.1038/s41586-019-1789-8}, \href {https://ui.adsabs.harvard.edu/abs/2019Natur.576...61G} {576, 61}

\bibitem[\protect\citeauthoryear{{Gentile Fusillo}, {G{\"a}nsicke}, {Farihi}, {Koester}, {Schreiber}  \& {Pala}}{{Gentile Fusillo} et~al.}{2017}]{fusilloetal17-1}
{Gentile Fusillo} N.~P.,  {G{\"a}nsicke} B.~T.,  {Farihi} J.,  {Koester} D.,  {Schreiber} M.~R.,   {Pala} A.~F.,  2017, \mn@doi [MNRAS] {10.1093/mnras/stx468}, \href {http://adsabs.harvard.edu/abs/2017MNRAS.468..971G} {468, 971}

\bibitem[\protect\citeauthoryear{{Gentile Fusillo} et~al.,}{{Gentile Fusillo} et~al.}{2019}]{fusilloetal19-1}
{Gentile Fusillo} N.~P.,  et~al., 2019, \mn@doi [\mnras] {10.1093/mnras/sty3016}, \href {https://ui.adsabs.harvard.edu/abs/2019MNRAS.482.4570G} {482, 4570}

\bibitem[\protect\citeauthoryear{{Gentile Fusillo} et~al.,}{{Gentile Fusillo} et~al.}{2021}]{fusilloetal21-1}
{Gentile Fusillo} N.~P.,  et~al., 2021, \mn@doi [\mnras] {10.1093/mnras/stab2672}, \href {https://ui.adsabs.harvard.edu/abs/2021MNRAS.508.3877G} {508, 3877}

\bibitem[\protect\citeauthoryear{{Guidry} et~al.,}{{Guidry} et~al.}{2021}]{guidryetal21-1}
{Guidry} J.~A.,  et~al., 2021, \mn@doi [\apj] {10.3847/1538-4357/abee68}, \href {https://ui.adsabs.harvard.edu/abs/2021ApJ...912..125G} {912, 125}

\bibitem[\protect\citeauthoryear{{Gunn} et~al.,}{{Gunn} et~al.}{2006}]{gunnetal06-1}
{Gunn} J.~E.,  et~al., 2006, \mn@doi [AJ] {10.1086/500975}, \href {http://adsabs.harvard.edu/abs/2006AJ....131.2332G} {131, 2332}

\bibitem[\protect\citeauthoryear{{Guy} et~al.,}{{Guy} et~al.}{2023}]{guyetal23-1}
{Guy} J.,  et~al., 2023, \mn@doi [\aj] {10.3847/1538-3881/acb212}, \href {https://ui.adsabs.harvard.edu/abs/2023AJ....165..144G} {165, 144}

\bibitem[\protect\citeauthoryear{{Hollands}, {Koester}, {Alekseev}, {Herbert}  \& {G{\"a}nsicke}}{{Hollands} et~al.}{2017}]{hollandsetal17-1}
{Hollands} M.~A.,  {Koester} D.,  {Alekseev} V.,  {Herbert} E.~L.,   {G{\"a}nsicke} B.~T.,  2017, \mn@doi [MNRAS] {10.1093/mnras/stx250}, \href {http://adsabs.harvard.edu/abs/2017MNRAS.467.4970H} {467, 4970}

\bibitem[\protect\citeauthoryear{{Hollands}, {G{\"a}nsicke}  \& {Koester}}{{Hollands} et~al.}{2018a}]{hollandsetal18-1}
{Hollands} M.~A.,  {G{\"a}nsicke} B.~T.,   {Koester} D.,  2018a, \mn@doi [\mnras] {10.1093/mnras/sty592}, \href {https://ui.adsabs.harvard.edu/abs/2018MNRAS.477...93H} {477, 93}

\bibitem[\protect\citeauthoryear{{Hollands}, {Tremblay}, {G{\"a}nsicke}, {Gentile-Fusillo}  \& {Toonen}}{{Hollands} et~al.}{2018b}]{hollandsetal18-2}
{Hollands} M.~A.,  {Tremblay} P.~E.,  {G{\"a}nsicke} B.~T.,  {Gentile-Fusillo} N.~P.,   {Toonen} S.,  2018b, \mn@doi [\mnras] {10.1093/mnras/sty2057}, \href {https://ui.adsabs.harvard.edu/abs/2018MNRAS.480.3942H} {480, 3942}

\bibitem[\protect\citeauthoryear{{Hollands}, {Tremblay}, {G{\"a}nsicke}  \& {Koester}}{{Hollands} et~al.}{2022}]{hollandsetal22-1}
{Hollands} M.~A.,  {Tremblay} P.~E.,  {G{\"a}nsicke} B.~T.,   {Koester} D.,  2022, \mn@doi [\mnras] {10.1093/mnras/stab3696}, \href {https://ui.adsabs.harvard.edu/abs/2022MNRAS.511...71H} {511, 71}

\bibitem[\protect\citeauthoryear{{Iben}, {Ritossa}  \& {Garcia-Berro}}{{Iben} et~al.}{1997}]{ibenetal97-1}
{Iben} I.~J.,  {Ritossa} C.,   {Garcia-Berro} E.,  1997, \mn@doi [ApJ] {10.1086/304822}, \href {1997ApJ...489..772I} {489, 772}

\bibitem[\protect\citeauthoryear{{Johnson}, {Aller}, {Howard}  \& {Crepp}}{{Johnson} et~al.}{2010}]{johnsonetal10-1}
{Johnson} J.~A.,  {Aller} K.~M.,  {Howard} A.~W.,   {Crepp} J.~R.,  2010, \mn@doi [PASP] {10.1086/655775}, \href {2010PASP..122..905J} {122, 905}

\bibitem[\protect\citeauthoryear{{Johnson}, {Klein}, {Koester}, {Melis}, {Zuckerman}  \& {Jura}}{{Johnson} et~al.}{2022}]{johnsonetal22-1}
{Johnson} T.~M.,  {Klein} B.~L.,  {Koester} D.,  {Melis} C.,  {Zuckerman} B.,   {Jura} M.,  2022, \mn@doi [\apj] {10.3847/1538-4357/aca089}, \href {https://ui.adsabs.harvard.edu/abs/2022ApJ...941..113J} {941, 113}

\bibitem[\protect\citeauthoryear{{Jura}}{{Jura}}{2003}]{jura03-1}
{Jura} M.,  2003, \mn@doi [ApJ Lett.] {10.1086/374036}, \href {2003ApJ...584L..91J} {584, L91}

\bibitem[\protect\citeauthoryear{{Kawka}, {Ferrario}  \& {Vennes}}{{Kawka} et~al.}{2023}]{kawkaetal23-1}
{Kawka} A.,  {Ferrario} L.,   {Vennes} S.,  2023, \mn@doi [\mnras] {10.1093/mnras/stad553}, \href {https://ui.adsabs.harvard.edu/abs/2023MNRAS.520.6299K} {520, 6299}

\bibitem[\protect\citeauthoryear{{Kilic}, {Bergeron}, {Kosakowski}, {Brown}, {Ag{\"u}eros}  \& {Blouin}}{{Kilic} et~al.}{2020}]{kilicetal20-1}
{Kilic} M.,  {Bergeron} P.,  {Kosakowski} A.,  {Brown} W.~R.,  {Ag{\"u}eros} M.~A.,   {Blouin} S.,  2020, \mn@doi [\apj] {10.3847/1538-4357/ab9b8d}, \href {https://ui.adsabs.harvard.edu/abs/2020ApJ...898...84K} {898, 84}

\bibitem[\protect\citeauthoryear{{Kleinman} et~al.,}{{Kleinman} et~al.}{2004}]{kleinmanetal04-1}
{Kleinman} S.~J.,  et~al., 2004, ApJ, \href {2004ApJ...607..426K} {607, 426}

\bibitem[\protect\citeauthoryear{{Koester}}{{Koester}}{2009}]{koester09-1}
{Koester} D.,  2009, \mn@doi [A\&A] {10.1051/0004-6361/200811468}, \href {2009A&A...498..517K} {498, 517}

\bibitem[\protect\citeauthoryear{{Koester} \& {Kepler}}{{Koester} \& {Kepler}}{2015}]{koester+kepler15-1}
{Koester} D.,  {Kepler} S.~O.,  2015, \mn@doi [A\&A] {10.1051/0004-6361/201527169}, \href {http://adsabs.harvard.edu/abs/2015A%26A...583A..86K} {583, A86}

\bibitem[\protect\citeauthoryear{{Koester}, {G{\"a}nsicke}  \& {Farihi}}{{Koester} et~al.}{2014}]{koesteretal14-1}
{Koester} D.,  {G{\"a}nsicke} B.~T.,   {Farihi} J.,  2014, \mn@doi [A\&A] {10.1051/0004-6361/201423691}, \href {2014A&A...566A..34K} {566, A34}

\bibitem[\protect\citeauthoryear{{Lagos}, {Schreiber}, {Zorotovic}, {G{\"a}nsicke}, {Ronco}  \& {Hamers}}{{Lagos} et~al.}{2021}]{lagosetal21-1}
{Lagos} F.,  {Schreiber} M.~R.,  {Zorotovic} M.,  {G{\"a}nsicke} B.~T.,  {Ronco} M.~P.,   {Hamers} A.~S.,  2021, \mn@doi [\mnras] {10.1093/mnras/staa3703}, \href {https://ui.adsabs.harvard.edu/abs/2021MNRAS.501..676L} {501, 676}

\bibitem[\protect\citeauthoryear{{Manser}, {G{\"a}nsicke}, {Gentile Fusillo}, {Ashley}, {Breedt}, {Hollands}, {Izquierdo}  \& {Pelisoli}}{{Manser} et~al.}{2020}]{manseretal20-1}
{Manser} C.~J.,  {G{\"a}nsicke} B.~T.,  {Gentile Fusillo} N.~P.,  {Ashley} R.,  {Breedt} E.,  {Hollands} M.,  {Izquierdo} P.,   {Pelisoli} I.,  2020, \mn@doi [\mnras] {10.1093/mnras/staa359}, \href {https://ui.adsabs.harvard.edu/abs/2020MNRAS.493.2127M} {493, 2127}

\bibitem[\protect\citeauthoryear{{Manser} et~al.,}{{Manser} et~al.}{2023}]{manseretal23-1}
{Manser} C.~J.,  et~al., 2023, \mn@doi [\mnras] {10.1093/mnras/stad727}, \href {https://ui.adsabs.harvard.edu/abs/2023MNRAS.521.4976M} {521, 4976}

\bibitem[\protect\citeauthoryear{{Melis}, {Klein}, {Doyle}, {Weinberger}, {Zuckerman}  \& {Dufour}}{{Melis} et~al.}{2020}]{melisetal20-1}
{Melis} C.,  {Klein} B.,  {Doyle} A.~E.,  {Weinberger} A.,  {Zuckerman} B.,   {Dufour} P.,  2020, \mn@doi [\apj] {10.3847/1538-4357/abbdfa}, \href {https://ui.adsabs.harvard.edu/abs/2020ApJ...905...56M} {905, 56}

\bibitem[\protect\citeauthoryear{{Mustill}, {Veras}  \& {Villaver}}{{Mustill} et~al.}{2014}]{mustilletal14-1}
{Mustill} A.~J.,  {Veras} D.,   {Villaver} E.,  2014, \mn@doi [MNRAS] {10.1093/mnras/stt1973}, \href {2014MNRAS.437.1404M} {437, 1404}

\bibitem[\protect\citeauthoryear{{Pelletier}, {Fontaine}, {Wesemael}, {Michaud}  \& {Wegner}}{{Pelletier} et~al.}{1986}]{pelletieretal86-1}
{Pelletier} C.,  {Fontaine} G.,  {Wesemael} F.,  {Michaud} G.,   {Wegner} G.,  1986, \mn@doi [ApJ] {10.1086/164410}, \href {1986ApJ...307..242P} {307, 242}

\bibitem[\protect\citeauthoryear{{Poleski} et~al.,}{{Poleski} et~al.}{2021}]{poleskietal21-1}
{Poleski} R.,  et~al., 2021, \mn@doi [\actaa] {10.32023/0001-5237/71.1.1}, \href {https://ui.adsabs.harvard.edu/abs/2021AcA....71....1P} {71, 1}

\bibitem[\protect\citeauthoryear{{Rocchetto}, {Farihi}, {G{\"a}nsicke}  \& {Bergfors}}{{Rocchetto} et~al.}{2015}]{rocchettoetal15-1}
{Rocchetto} M.,  {Farihi} J.,  {G{\"a}nsicke} B.~T.,   {Bergfors} C.,  2015, \mn@doi [MNRAS] {10.1093/mnras/stv282}, \href {http://adsabs.harvard.edu/abs/2015MNRAS.449..574R} {449, 574}

\bibitem[\protect\citeauthoryear{{Schreiber}, {G{\"a}nsicke}, {Toloza}, {Hernandez}  \& {Lagos}}{{Schreiber} et~al.}{2019}]{schreiberetal19-1}
{Schreiber} M.~R.,  {G{\"a}nsicke} B.~T.,  {Toloza} O.,  {Hernandez} M.-S.,   {Lagos} F.,  2019, \mn@doi [\apjl] {10.3847/2041-8213/ab42e2}, \href {https://ui.adsabs.harvard.edu/abs/2019ApJ...887L...4S} {887, L4}

\bibitem[\protect\citeauthoryear{{Trierweiler}, {Doyle}  \& {Young}}{{Trierweiler} et~al.}{2023}]{trierweileretal23-1}
{Trierweiler} I.~L.,  {Doyle} A.~E.,   {Young} E.~D.,  2023, \mn@doi [arXiv e-prints] {10.48550/arXiv.2306.03743}, \href {https://ui.adsabs.harvard.edu/abs/2023arXiv230603743T} {p. arXiv:2306.03743}

\bibitem[\protect\citeauthoryear{{Vanderbosch} et~al.,}{{Vanderbosch} et~al.}{2020}]{vanderboschetal19-1}
{Vanderbosch} Z.,  et~al., 2020, \mn@doi [\apj] {10.3847/1538-4357/ab9649}, \href {https://ui.adsabs.harvard.edu/abs/2020ApJ...897..171V} {897, 171}

\bibitem[\protect\citeauthoryear{{Vanderbosch} et~al.,}{{Vanderbosch} et~al.}{2021}]{vanderboschetal21-1}
{Vanderbosch} Z.~P.,  et~al., 2021, arXiv e-prints, \href {https://ui.adsabs.harvard.edu/abs/2021arXiv210602659V} {p. arXiv:2106.02659}

\bibitem[\protect\citeauthoryear{{Vanderburg} et~al.,}{{Vanderburg} et~al.}{2015}]{vanderburgetal15-1}
{Vanderburg} A.,  et~al., 2015, \mn@doi [Nat] {10.1038/nature15527}, \href {http://adsabs.harvard.edu/abs/2015Natur.526..546V} {526, 546}

\bibitem[\protect\citeauthoryear{{Vanderburg} et~al.,}{{Vanderburg} et~al.}{2020}]{vanderburgetal20-1}
{Vanderburg} A.,  et~al., 2020, \mn@doi [\nat] {10.1038/s41586-020-2713-y}, \href {https://ui.adsabs.harvard.edu/abs/2020Natur.585..363V} {585, 363}

\bibitem[\protect\citeauthoryear{{Veras} \& {G{\"a}nsicke}}{{Veras} \& {G{\"a}nsicke}}{2015}]{veras+gaensicke15-1}
{Veras} D.,  {G{\"a}nsicke} B.~T.,  2015, \mn@doi [MNRAS] {10.1093/mnras/stu2475}, \href {http://adsabs.harvard.edu/abs/2015MNRAS.447.1049V} {447, 1049}

\bibitem[\protect\citeauthoryear{{Veras} \& {Hinkley}}{{Veras} \& {Hinkley}}{2021}]{veras+hinkley21-1}
{Veras} D.,  {Hinkley} S.,  2021, \mn@doi [\mnras] {10.1093/mnras/stab1311}, \href {https://ui.adsabs.harvard.edu/abs/2021MNRAS.505.1557V} {505, 1557}

\bibitem[\protect\citeauthoryear{{Villaver} \& {Livio}}{{Villaver} \& {Livio}}{2009}]{villaver+livio09-1}
{Villaver} E.,  {Livio} M.,  2009, \mn@doi [ApJ Lett.] {10.1088/0004-637X/705/1/L81}, \href {2009ApJ...705L..81V} {705, L81}

\bibitem[\protect\citeauthoryear{{Wilson}, {Farihi}, {G{\"a}nsicke}  \& {Swan}}{{Wilson} et~al.}{2019}]{wilsonetal19-1}
{Wilson} T.~G.,  {Farihi} J.,  {G{\"a}nsicke} B.~T.,   {Swan} A.,  2019, \mn@doi [\mnras] {10.1093/mnras/stz1050}, \href {https://ui.adsabs.harvard.edu/abs/2019MNRAS.tmp.1000W} {p.~1000}

\bibitem[\protect\citeauthoryear{{Xu}, {Zuckerman}, {Dufour}, {Young}, {Klein}  \& {Jura}}{{Xu} et~al.}{2017}]{xuetal17-1}
{Xu} S.,  {Zuckerman} B.,  {Dufour} P.,  {Young} E.~D.,  {Klein} B.,   {Jura} M.,  2017, \mn@doi [ApJ Lett.] {10.3847/2041-8213/836/1/L7}, \href {http://adsabs.harvard.edu/abs/2017ApJ...836L...7X} {836, L7}

\bibitem[\protect\citeauthoryear{{Zhu} \& {Dong}}{{Zhu} \& {Dong}}{2021}]{zhu+dong21-1}
{Zhu} W.,  {Dong} S.,  2021, \mn@doi [\araa] {10.1146/annurev-astro-112420-020055}, \href {https://ui.adsabs.harvard.edu/abs/2021ARA&A..59..291Z} {59, 291}

\bibitem[\protect\citeauthoryear{{Zhu}, {Petrovich}, {Wu}, {Dong}  \& {Xie}}{{Zhu} et~al.}{2018}]{zhuetal18-1}
{Zhu} W.,  {Petrovich} C.,  {Wu} Y.,  {Dong} S.,   {Xie} J.,  2018, \mn@doi [\apj] {10.3847/1538-4357/aac6d5}, \href {https://ui.adsabs.harvard.edu/abs/2018ApJ...860..101Z} {860, 101}

\bibitem[\protect\citeauthoryear{{Zuckerman} \& {Becklin}}{{Zuckerman} \& {Becklin}}{1987}]{zuckerman+becklin87-1}
{Zuckerman} B.,  {Becklin} E.~E.,  1987, \mn@doi [Nat] {10.1038/330138a0}, \href {1987Natur.330..138Z} {330, 138}

\bibitem[\protect\citeauthoryear{{Zuckerman}, {Koester}, {Reid}  \& {H{\"u}nsch}}{{Zuckerman} et~al.}{2003}]{zuckermanetal03-1}
{Zuckerman} B.,  {Koester} D.,  {Reid} I.~N.,   {H{\"u}nsch} M.,  2003, \mn@doi [ApJ] {10.1086/377492}, \href {2003ApJ...596..477Z} {596, 477}

\bibitem[\protect\citeauthoryear{{Zuckerman}, {Koester}, {Melis}, {Hansen}  \& {Jura}}{{Zuckerman} et~al.}{2007}]{zuckermanetal07-1}
{Zuckerman} B.,  {Koester} D.,  {Melis} C.,  {Hansen} B.~M.,   {Jura} M.,  2007, \mn@doi [ApJ] {10.1086/522223}, \href {2007ApJ...671..872Z} {671, 872}

\bibitem[\protect\citeauthoryear{{Zuckerman}, {Melis}, {Klein}, {Koester}  \& {Jura}}{{Zuckerman} et~al.}{2010}]{zuckermanetal10-1}
{Zuckerman} B.,  {Melis} C.,  {Klein} B.,  {Koester} D.,   {Jura} M.,  2010, \mn@doi [ApJ] {10.1088/0004-637X/722/1/725}, \href {2010ApJ...722..725Z} {722, 725}

\makeatother
\end{thebibliography}







\bsp	
\label{lastpage}
\end{document}